\documentclass[prd,aps,nofootinbib,preprint]{revtex4}
\usepackage{amsmath,amssymb,amsfonts,color,graphicx,graphics,latexsym,placeins,epsfig,multirow}
\usepackage{array}
\newcolumntype{P}[1]{>{\centering\arraybackslash}p{#1}}
\usepackage{footmisc}
\usepackage[compatibility=false]{caption}
\usepackage{subcaption}
\usepackage{comment}
\usepackage{bbold}
\usepackage{enumitem}
\setdescription{leftmargin=12.5pt}
\usepackage{bbold}
\usepackage{hyperref}
\DeclareMathOperator{\sech}{sech}
\usepackage{natbib}
\setcitestyle{square, comma, numbers, sort&compress}
\usepackage{float}
\newcommand*{\id}{{\rm\hbox{1\kern-0.15em \vrule width .1pt depth-.2pt}}}
\newcommand{\noi}{\noindent}

\begin{document}

%\preprint{LIGO-P1800035}

\title{\Large \bf Kundt wave geometries in Eddington-inspired Born-Infeld gravity: New solutions and memory effects}
\author{Indranil Chakraborty}
\email{indradeb@iitkgp.ac.in}
\affiliation{Department of Physics \\ Indian Institute of Technology Kharagpur, 721 302, India}
%\affiliation{${}^2$ Centre for Theoretical Studies \\ Indian Institute of Technology Kharagpur, 721 302, India}

\begin{abstract}
\noi We explore memory effects for novel Kundt wave spacetimes in the recently proposed Eddington-inspired Born-Infeld (EiBI) theory of gravity. First, we construct new, exact Kundt wave geometries in this theory for two different  matter sources \textemdash (i) generic matter designed to satisfy the field equations as well as energy conditions, and (ii) electromagnetic field. For both sources we find that the EiBI theory parameter $\kappa$ couples only with the nonradiative part of the physical metric solution. Thereafter, we solve the geodesic and the geodesic deviation equations, in the above spacetimes with the aim of arriving at memory effects. This analysis is carried out numerically and reveals unique memory features  depending on the type of matter source present and the signature of the spacetime scalar curvature. The role of $\kappa$ in influencing the memory effect, for a given background spacetime, is also noted. Thus, apart from providing novel radiative solutions in EiBI gravity, we also show how different matter configurations are responsible for distinct memory characteristics.

 % \noi The predictions of the theory significantly deviates from GR at large curvatures and high densities. 

\end{abstract}

\maketitle
new

\section{Introduction}

\noi A major achievement of gravitational wave astronomy \citep{Abbott:2016} has been its role in probing the strong-gravity regime \citep{Berti:2018}.  The gravitational wave memory effect is one of the few unobserved predictions of strong-field gravity that remains elusive to date \citep{Favata:2010}. The memory effect causes permanent change in the relative position (or relative velocity) of the freely falling detectors after the passage of a gravitational wave pulse. It is predicted to yield a net constant shift in the overall gravitational wave amplitude \citep{Favata:2010,bieri2015gravitational}.

\noi %Investigations on memory effects have risen significantly, of late, as it lends support to both theoretical and phenomenological aspects of gravitational physics. On the theoretical side, detection of gravitational memory would possibly point to the presence of soft gravitons, following from the infrared triangle conjecture by Strominger \citep{Strominger:2016, Strominger:2017}. It would also establish the infinite dimensional BMS group (named after Bondi, Metzner and Sachs) \citep{Bondi:1962, Sachs:1962} as the  asymptotic symmetry group for isolated sources in asymptotically flat spacetimes. These links have also been extended to other gauge theories like electromagnetism \cite{Pasterski:2017} and QCD \cite{Pate:2017}. Moreover, the infinite dimensional BMS group generates an infinite number of conservation equations. Any candidate waveform, obtained using numerical relativity simulations or effective-one-body approach for a binary merger, has to satisfy these conservation laws (also called {\em BMS flux-balance laws}) \citep{Ashtekar:2019,Khera:2021}. Thus, extracting gravitational memory for these test waveforms can act as a cross-check of their accuracy. Such phenomenological studies have been worked out in \citep{Mitman:2020} for binary black hole mergers. 

\noi Earlier work on memory effects used the linearized gravitational theory to predict a permanent change in the geodesic separation after the pulse has departed \citep{Zeldovich:1974,Braginsky:1985}. This is caused due to the change in the double derivative of the quadrupole moment of the source before and after the burst of radiation \cite{Braginsky:1987}.  Later, Christodoulou showed that the stress energy of gravitational waves reaching null infinity can also contribute to the memory effect \citep{Christodoulou:1991}. These two types of memories were subsequently  named in the literature as the ordinary (linear) and null (nonlinear) memory effects \citep{Bieri:2013inv, Tolish:2014, Madler:2016}.  {Currently, there are ongoing efforts to detect this effect in ground-based observatories \citep{Boersma:2020,Hubner:2021}, space-based detectors \citep{Burko:2020} and pulsar timing arrays \citep{Aggarwal:2019}. {Apart from being an astronomical observable, memory effects provide connection with different aspects of theoretical physics like soft theorems and asymptotic symmetries \citep{Strominger:2016,Strominger:2017}. }}

\noi { The study of memory effects can also be formulated in {\em exact solutions of spacetimes containing gravitational waves} \citep{Zhang:2017,Zhang:2017soft,Chakra:2020,Srijit:2019,Flanagan:2019}. In \citep{Zhang:2017soft}, Zhang et al. studied geodesic evolution in vacuum exact plane waves by considering a Gaussian pulse profile. Numerically integrating the geodesic equations they showed how gravitational wave memory effects are realized in such geometries. Building on these ideas, we earlier worked out memory effects in Kundt wave geometries for General Relativity (GR) \citep{Chak1:2020} and Brans-Dicke (BD) \citep{Siddhant:2020} theory. In this article, we extend this analysis to a recently proposed modified theory called Eddington-inspired Born-Infeld (EiBI) gravity.}

\noi In general, Kundt spacetimes are defined as having a null geodesic congruence where all the optical scalars vanish \citep{Kundt:1961, Stephani:2003,Griffiths:2009}. The tangent vector to the congruence is, in general, not covariantly constant \citep{Griffiths:2009}.  Such spacetimes admit gyratons which are spinning null sources and generate angular momentum in the spacetime \citep{Frolov:2005, Frolov1:2005,Podolsky:2008,Kadlecova:2009,Krtous:2012,Podolsky:2014,Podolsky_gyr:2019}. Apart from GR \citep {Podolsky:1998,Bicak1:1999,Bicak:1999,Podolski:2001,Ortaggio:2002,Griffiths:2003,Coley:2009,Svarc:2012,Podolski:2013,Tahamtan:2017,Ortaggio:2018} and  BD gravity \citep{Siddhant:2020}, theories like quadratic gravity \citep{Pravda:2017,Svarc_Quadgrav:2018,Podolsky_Quadgrav:2019}, Gauss-Bonnet theory \citep{Svarc:2020} and Infinite Derivative gravity \citep{Kolar:2021} have been investigated while studying this geometry.  {For a brief review on Kundt spacetimes we refer the reader to \citep{Coley:2009}.  Kundt wave spacetimes are generalizations of exact plane waves propagating in a curved background spacetime. The presence of matter is responsible for the nonplanarity of the wavefronts.
On the contrary, exact plane waves have planar wavefronts and flat background geometry. Any solution obtained for exact plane waves in EiBI gravity will be same with GR since the two theories are identical in vacuum. Hence, we work with Kundt waves and try to examine how the matter content present in this geometry helps to bring out features of EiBI theory.}

\noi {There exist nontrivialities in obtaining memory effects for spacetimes having nonzero background curvature. Pursuing a technique formulated in \citep{Chu:2019}, we showed in our earlier work on BD gravity \citep{Siddhant:2020} how gravitational memory contribution can be isolated from the background by solving the geodesic deviation equation.}  Also, we found that the results obtained by solving geodesic equations qualitatively match with those found from the total deviation (wave and background). The deviation due to gravitational wave ({\em memory effect}) is not completely obtained by studying the geodesic equations only. Hence, we work out the memory effects here, using both geodesic equations and geodesic deviation equations.

\noi In the present article, we choose two different types of matter sources and look for the nature of the memory effects they exhibit. The first one is a generic source. It is designed to satisfy all the energy conditions and field equations. The next one is the electromagnetic (EM) field. We find exact solutions in both these cases. After constructing the metric solution, we study their memory effects. At first, we perform geodesic analysis and then we analyze memory effects using the geodesic deviation. Gravitational memory effects are shown to be dependent on the EiBI parameter $\kappa$ and the choice of the matter source. 

\noi The article is organized as follows. In Sec. II, we provide an overview of the basic framework used in the paper. Section II A deals with a brief recap of EiBI gravity. Section II B focuses on Kundt wave geometry.  {In Sec. II C, we give a brief summary of how memory effects are realized in related physical scenarios and provide the requisite connection with our present work.} Moreover, the methodology used to calculate memory effects using geodesic equations and the deviation equation is also given. The entire deviation equation formalism used here can also be found in \citep{Siddhant:2020}. In Sec. III, we present the exact solution and memory effects for the generic matter source. Section IV deals with the EM field as a source. Finally, we conclude in Sec. V with comments on future work. An appendix is provided at the end listing the Riemann tensors in the tetrad frame used in the geodesic deviation analysis.
 
\section{Basic framework}

\subsection{Eddington-inspired Born-Infeld gravity: A brief recap}

\noi A determinant based action for gravitational theories was first introduced by Eddington \citep{Eddington:1930} by taking a pure connection dependent Lagrangian. His theory was identical to GR for vacuum constant curvature spacetimes.  Around the same time, in electrodynamics, Born and Infeld \citep{Born:1934}  regularized the field divergences at the source of a point charge by also introducing a determinantal form of the action. Several years later, a gravitational analog of such a theory was proposed by Deser and Gibbons in \citep{Deser:1998}. Being a pure metric theory, it suffered from unconstrained higher derivative curvature terms. 

\noi Combining these ideas, Vollick \citep{Vollick:2004,Vollick:2005} used the Palatini approach, where both geometry and matter were coupled to the metric and the connection. In our work, we focus on a theory recently proposed by Banados and Ferreira \citep{Banados:2010}, where the matter coupling is simpler and is only dependent on the metric. The theory has subsequently been called Eddington-inspired Born-Infeld (EiBI) gravity. The action of EiBI gravity is,
\begin{equation}
    S_{EiBI}(g,\Gamma,\Psi)=\frac{2}{\kappa}\int d^4x\bigg[\sqrt{-|g_{\mu\nu}+\kappa R_{\mu\nu}(\Gamma)|}-(1+\kappa\Lambda)\sqrt{-|g_{\mu\nu}|}\bigg] + S_M(g,\Psi). \label{eq:EiBI_action}
\end{equation}

\noi $\kappa$ is the only parameter of EiBI theory. It has dimensions inverse of the cosmological constant. In Vollick's action \citep{Vollick:2005}, the matter coupling occurs inside the determinant. The action in Eq.(\ref{eq:EiBI_action}) reduces to the Einstein-Hilbert action in two ways. First, for a vacuum solution ($S_M=0$). Second, in the GR limit, $g_{\mu\nu}>>\kappa R_{\mu\nu}$. Thus, the difference between EiBI gravity and GR occurs for regions involving nonzero matter and higher curvatures.

\noi From the previous discussion, it is obvious that GR and EiBI gravity differ significantly in the strong-field regime. Hence, EiBI theory can be used as a test bed to study the physics of the early universe. In fact, Banados and Ferreira, in their original work \citep{Banados:2010}, showed that EiBI cosmology predicts a maximum density of the universe, successfully evading the big bang singularity. Most of the literature on EiBI gravity is focused on the phenomenological implications it offers for astrophysics and cosmology \citep{Pani:2011,Escamilla:2012,Jana:2015,Shaikh:2015,Jana:2017,Feng:2018}. We refer the reader to the review \citep{Beltran:2017} for more recent works on EiBI gravity. 

\noi Coming back to the action given in Eq.(\ref{eq:EiBI_action}), we find that the metric and the connection are considered independent to each other. The variation w. r. t. the connection ($\Gamma$) gives the condition of metric compatibility\footnote{$\tilde\nabla_\gamma (\sqrt{-|q_{\mu\nu}|}q^{\mu\nu})=0$} of $q_{\mu\nu}$, which is defined as
\begin{equation}
    q_{\mu\nu}=g_{\mu\nu}+\kappa R_{\mu\nu}(\Gamma) \label{eq:field_eqn_1}
    \end{equation}
    
\noi $g_{\mu\nu}$ and $q_{\mu\nu}$ are termed as the physical metric and the auxiliary metric, respectively. Since, $q_{\mu\nu}$ is compatible with the covariant derivative associated with the connection, the Ricci tensor appearing in the R. H. S. of Eq.(\ref{eq:field_eqn_1}) can be constructed from the auxiliary metric. The variation w. r. t. $g_{\mu\nu}$ yields,
\begin{equation}
    \sqrt{-|q_{\mu\nu}|}q^{\mu\nu}=(1+\kappa\Lambda)\sqrt{-|g_{\mu\nu}|}g^{\mu\nu}-\kappa\sqrt{-|g_{\mu\nu}|}T^{\mu\nu} \label{eq:field_eqn_2}
\end{equation}

\noi We also note that EiBI theory is a bimetric theory of gravity \citep{Isham:1971} as $q_{\mu\nu}$ does not couple with matter. Like any local gravitational theory, the energy-momentum tensor $T_{\mu\nu}$ satisfies the conservation equation. The zero value of the covariant divergence of $T_{\mu\nu}$ is compatible with the physical metric.

\noi As stated earlier, here we investigate memory effects in Kundt wave geometries in EiBI theory. At first, we construct exact solutions for two different types of matter sources. Then, we perform geodesic analysis and geodesic deviation analysis to arrive at the memory effects. Our primary aim is to find how the features  of  memory effects change for different matter sources. We also try to find how the variation of $\kappa$ influences the nature of memory. In our analysis, we do not restrict the value of $\kappa$.

\subsection{Kundt wave geometry}

\noi The general Kundt spacetime line element in Brinkmann coordnates ($u,v,x,y$) is \citep{Stephani:2003, Griffiths:2009},
\begin{equation}
    ds^2=-H(u,v,x,y)du^2-2dudv-2W_1(u,v,x,y)dudx-2W_2(u,v,x,y)dudy+\frac{dx^2+dy^2}{P(u,x,y)^2}.
\end{equation}

\vspace{0.1in}

\noi The null vector $n^\mu=\delta^\mu\,_v$ is normal to the transverse spatial plane spanned by the tangent vectors $P\partial_x$ and $P\partial_y$. Along $n^\mu$, the null congruence has vanishing optical scalars. We work with a special type of Kundt geometry known as {\em Kundt waves}. The spacetime metric takes the form 
\begin{equation}
    ds^2=-H(u,x,y)du^2-2dudv+\frac{dx^2+dy^2}{P(u,x,y)^2} \label{eq:Kwave_general}
\end{equation}

\noi The term $P(u,x,y)$ provides the relevant background curvature (nonradiative) while the $g_{uu}$ component ($H(u,x,y)$) of the metric gives us the gravitational wave contribution. One can get back the well-known $pp$-wave solutions by setting $P=1$, {\em i.e.} the wave propagates in the Minkowski (flat) background. Wavefronts ($u$-constant hypersurfaces) for a Kundt wave geometry are curved due to the presence of matter/cosmological constant in GR. For Brans-Dicke gravity, we obtained a vacuum solution \citep{Siddhant:2020} where the scalar field provides an {\em effective matter} contribution.

\noi We consider two kinds of matter sources. The first one is a generic ({\em i.e.} without any explicit matter Lagrangian) matter source and the second one is the well-known Maxwell EM field. As mentioned earlier, the presence of matter is required so that the solution in EiBI theory differs from GR.

\subsection{Gravitational memory effect}

\noi  Memory effect is basically a measure of the  gravitational wave deviation. Thus, one can study it either by solving the geodesic deviation equations \citep{Braginsky:1985} or observing the evolution of the separation between a pair of geodesic trajectories by evaluating the geodesic equations \citep{Zhang:2017, Zhang:2017soft}. Gravitational memory is easily realized in linearized gravity by the following equation. 
\begin{equation}
\Delta\xi^{i}=\frac{1}{2}\Delta(h^i\,_j)^{TT}\xi^j \label{eq:TT_gauge}
\end{equation}

\noi Generally for a gravitational wave burst scenario, we find that the metric perturbation in the transverse traceless gauge ($h_{ij}^{TT}$) differs at early and late times \citep{Zeldovich:1974}. Hence, the change in proper length is nonzero ($\Delta\xi^{i}\neq 0$) and there is a displacement memory effect. Also, if $\Delta\xi^{i}$ varies at late times, we have a velocity memory effect \citep{Zhang_vel:2018}.  

\noi Finding out memory in exact solutions like plane gravitational waves was first done in  \citep{Zhang:2017,Zhang:2017soft}. Exact plane waves are nonlinear generalizations of the linearized plane gravitational wave metric in $TT$ gauge. This fact is made obvious by writing the metric line element in the Baldwin-Jeffrey-Rosen (BJR) coordinate system \citep{Rosen:1937}. The metric function denoting the radiation field $a_{ij}$ (look at Eq.(2.4) in \citep{Zhang:2017soft}) differs before and after the onset of the pulse. This signifies the presence of a memory effect in analogy with Eq.(\ref{eq:TT_gauge}). But BJR coordinates suffer from singularities and, hence, Brinkmann coordinates were used in finding out displacement and velocity memory effects.

\noi  We have continued to work in Brinkmann coordinates in the case of Kundt geometries. In GR \citep{Chak1:2020}, we only analyzed the geodesic equations. We numerically obtained distinct memory effects corresponding to positive and negative scalar curvature solutions. For BD gravity \citep{Siddhant:2020}, we obtained an exact solution. Thereafter, we studied memory effects for two different values of $\omega$ by solving geodesic equations and the deviation equations. An illustrative analytical solution was obtained for $\omega=-2$ which corresponds to the constant negative curvature solution. For the other case, $\omega = +1$, we numerically obtained different
memory effects as compared to the positive curvature solutions of GR. In our current work, we perform a similar analysis (constructing solutions and evaluating memory effects) as was done for the BD theory.

\noi Below, we discuss briefly the methods used to calculate memory effects here. We also comment on the differences in the methods applied.   

\subsubsection{Geodesic memory effect}

\noi Apart from plane waves and Kundt waves, finding memory effects using geodesics have been worked out for radiative spacetimes like gyratons \citep{Shore:2018} and gravitational shockwaves \citep{Shore:2018, Srijit:2019}. The general methodology to arrive at the memory effect is, pointwise, noted below. 

\noi \textbullet The gravitational wave term in the metric is chosen to be {\em pulselike}. 

\noi  \textbullet The geodesic equations are solved (analytically/ numerically) for a pair of geodesics (or more than two) having initial transverse coordinate velocity set to zero.

\noi \textbullet The change in the separation between the two geodesic trajectories, before and after the pulse, is noted.

\noi \textbullet This change in separation is termed as the {\em displacement memory effect}. If the separation is not constant, then differentiating the geodesic separation gives a measure of the {\em velocity memory effect}. 

\noi Here, we try to look at how the geodesic separation evolves for different values of the EiBI parameter $\kappa$.

\subsubsection{Gravitational memory using geodesic deviation}

\noi Geodesic deviation analysis can also be used to find out the gravitational memory effect \citep{Braginsky:1985}. Here we employ the same procedure as done in \citep{Siddhant:2020}, to find out the memory effects. This technique was first applied to calculate memory in AdS spacetime \citep{Chu:2019}. We briefly mention the salient features of this technique, pointwise, below.

\vspace{0.1in}

\noi \textbullet Fermi normal coordinates $(t,Z^i)$ are constructed along a chosen timelike geodesic such that the Christoffel connections vanish along that curve.

\noi \textbullet A parallel propagated tetrad ($e_i\,^\mu$) along that geodesic\footnote{Spacetime coordinates are denoted by Greek indices $(\mu,\nu,..)$ while the Fermi coordinates are given by Latin indices $(i,j,...)$.} is obtained in which the tangent to the geodesic curve is taken as $e_0\,^\mu$. 

\noi  \textbullet The geodesic deviation vector ($\xi^\mu$) in the coordinate basis is related to the deviation vector in the Fermi basis ($Z^i$), via
\begin{equation}
    \xi^\mu=Z^i e_i\,^\mu \label{eq:dev_tetrad}
\end{equation}

\noi \textbullet The deviation equation in the Fermi coordinate, then, reduces to a Jacobi equation.
\begin{equation}
    \frac{d^2Z^i}{dt^2}=-{R}^i\,_{0j0} Z^j \label{eq:deviation_eqn1}
\end{equation}

\noi  \textbullet The Riemann tensor in the tetrad frame is split into the background and wave. The wave contribution comes from the terms proportional to $H(u,x,y)$ (or its derivatives).

\noi \textbullet The respective deviation equations for the background and the wave become
\begin{gather}
    \frac{d^2Z^i_B}{dt^2}=-({R}^i\,_{0j0})_B Z^j_B \label{eq:deviation_eqn1_bg}\\
    \frac{d^2Z^i_W}{dt^2}=-[({R}^i\,_{0j0})_B+({R}^i\,_{0j0})_W] Z^j_W-({R}^i\,_{0j0})_W Z^j_B\label{eq:deviation_eqn1_wave}
\end{gather}

\noi \textbullet  Eqs.(\ref{eq:deviation_eqn1_bg}) and (\ref{eq:deviation_eqn1_wave}) are solved and the results obtained are transformed back to the coordinate basis using Eq.(\ref{eq:dev_tetrad}).

\noi \textbullet The total deviation  is obtained by adding the contributions of the background and the gravitational wave ($\xi^\mu=\xi^\mu_B+\xi^\mu_W$).

\vspace*{0.3cm}
\noi This entire deviation analysis can also be studied in the coordinate basis. We approach the problem using the Fermi basis to simplify the calculations. Unlike the geodesic equations, the deviation equation is perturbative in nature. Hence, the separation obtained from both the methods will not be exactly similar, but shall match {\em qualitatively}. Such a comparison was done in our earlier work \citep{Siddhant:2020} where we demonstrated that geodesic analysis is not sufficient to bring out, exclusively, the amount of gravitational memory (the
memory effect is measured by the gravitational wave deviation only).
This is because the geodesic deviation equation, being {\em linear}, can be split into its respective background and wave components. On the other hand, the geodesic equations are nonlinear and give a combined separation by having the contributions of both the background and wave. Thus, the results from the geodesic analysis were in agreement with the solutions of the total deviation. In our work here we will also look into similar features, in detail, within the present context.

\section{Kundt wave metric with a generic matter source}

\noi We start by writing down the ans{\"a}tze for the physical metric and auxiliary metric, respectively"
\begin{align}
 ds^2=g_{\mu\nu}dx^\mu dx^\nu=-H_1(u,x,y)du^2-2dudv+\frac{dx^2+dy^2}{P_1(x,y)^2} \label{eq:phys_metric_KW}   \\
  ds^2=q_{\mu\nu}dx^\mu dx^\nu=-H_2(u,x,y)du^2-2dudv+\frac{dx^2+dy^2}{P_2(x,y)^2} \label{eq:aux_metric_KW}
\end{align}

\noi We assume that the metric functions $P_1$ and $P_2$ are independent of $u$. The $uu$-component of Eq.(\ref{eq:field_eqn_1}) gives,
\begin{equation}
     -H_2=-H_1+\frac{\kappa}{2}P_2\,^2(H_2,_{xx}+H_2,_{yy}). \label{eq:uu_eqn_1}
\end{equation}

\noi Setting $H_2(u,x,y)=h(u)(x^2-y^2)$, we find, $H_1=H_2$. 
\begin{equation}
    H_1(u,x,y)=h(u)(x^2-y^2) \label{eq:H_1_soln}
\end{equation}
\noi This solution corresponds to the plus polarization of gravitational wave. The wave profile is given by the term $h(u)$. 

\noi From the $xx$ (or $yy$) component of  Eq.(\ref{eq:field_eqn_1}) we get,
 \begin{equation}
    \frac{1}{P_2\,^2}=\frac{1}{P_1\,^2}-\kappa\frac{P_2,_y^2+P_2,_x^2-P_2(P_2,_{xx}+P_2,_{yy})}{P_2\,^2}. \label{eq:xx_eqn_1}
\end{equation}

\noi Eqs.(\ref{eq:uu_eqn_1}) and (\ref{eq:xx_eqn_1}) are only dependent on the geometry of the theory. The field equation (\ref{eq:field_eqn_2}) requires specification of the matter content in the spacetime. In this section, we work with a generic source, which is well-suited for obtaining an exact solution. We will also check, how the relevant energy conditions behave, for this matter source. After finding the metric solution, we go over to the study of gravitational memory effects.

 \noi Analyzing solutions for such a generalized source can act as a template for {\em comparing memory effects with other known sources}. In this paper, we try to investigate these comparisons corresponding to the EM field.

\subsection{Exact solution}

\noi From Eq.(\ref{eq:field_eqn_2}), we find that only $uv, vv, xx$ (or $yy$) yield nontrivial equations.
\begin{equation}
    uv: \hspace{2cm} \frac{1}{P_2\,^2}=\frac{1+\kappa\Lambda}{P_1\,^2}+\frac{\kappa}{P_1\,^2}T^{uv} \label{eq:uv_eqn_2}
\end{equation}

\noi Eq.(\ref{eq:uv_eqn_2}) shows that $P_1$ and $P_2$ are conformally related to each other. In the case where $T^{uv}$ becomes a constant, they are related via scaling. 
\begin{equation}
    vv: \hspace{2cm} \frac{H_2}{P_2\,^2}=(1+\kappa\Lambda) \frac{H_1}{P_1\,^2}-\frac{\kappa}{P_1\,^2}T^{vv} \label{eq:vv_eqn_2}
\end{equation}

\noi Comparing Eq.(\ref{eq:vv_eqn_2}) with Eq.(\ref{eq:uv_eqn_2}), we find, $T^{vv}=-H_1T^{uv}$. The equation for the $xx$-component simply gives
\begin{equation}
    T^{xx}=\Lambda P_1\,^2 \label{eq:xx_eqn_2}
\end{equation}

\noi We find that there are five unknowns ($P_1, P_2, T^{uv}, T^{vv}, T^{xx}$) having four independent equations (\ref{eq:xx_eqn_1}-\ref{eq:xx_eqn_2}). We choose $T^{uv}=\sigma$ (constant) , and try to solve for the other unknowns. The components of $T^{\mu\nu}$ which vanish from the field equation (\ref{eq:field_eqn_2}) are identically taken equal to zero.
 
\noi In this class of metrics, $u$ acts as an {\em effective time} parameter.\footnote{It is the affine parameter for the metric line element in Eq.(\ref{eq:Kwave_general}). A dot overhead (as used in Eqs.(\ref{eq:x_eqn_KW1}), (\ref{eq:y_eqn_KW1}) etc.) means differentiation w. r. t. the affine parameter $u$.} Thus, the constant $\sigma$ can be attributed to a {\em matter flux} present in the spacetime.  Plugging $T^{uv}=\sigma$ into Eq.(\ref{eq:uv_eqn_2}) gives a scaling relation between $P_1$ and $P_2$,
\begin{equation}
P_1\,^2=P_2\,^2(1+\kappa(\Lambda+\sigma))  \label{eq:scaling}  
\end{equation}

\noi We find that if the flux is nondynamical, then the induced metrics on the wavefronts are related via scaling. Substituting Eq.(\ref{eq:scaling}) in Eq.(\ref{eq:xx_eqn_1}) gives,
    \begin{equation}
    P_2(P_2,_{xx}+P_2,_{yy})-P_2,_x^2-P_2,_y^2=\frac{\Lambda+\sigma}{1+\kappa(\Lambda+\sigma)} =\alpha \label{eq:P_2_eqn}
\end{equation}

\noi  Here, $\alpha$ is a constant. Solving Eq.(\ref{eq:P_2_eqn}) provides analytical forms of $P_2$ and $P_1$,
\begin{equation}
    P_2=1+\frac{\alpha}{4}(x^2+y^2) \hspace{1cm} P_1=\sqrt{1+\kappa(\Lambda+\sigma)} \bigg[1+\frac{\alpha}{4}(x^2+y^2)\bigg] \label{eq:metric_P1_P2}
\end{equation}
\noi After solving the field equations, we find that the pulse profile $h(u)$ in Eq.(\ref{eq:H_1_soln}) is unconstrained from the field equations. The  auxiliary metric is independent of $\kappa$. Eq.(\ref{eq:metric_P1_P2}) shows that $\kappa$ only couples to the physical metric via $P_1$ which is nonradiative. Even the matter coupling ($\sigma$) occurs through  $P_1$. Thus, the induced metric on the $u$-constant hypersurface is dependent on the underlying theory.  {If the analysis was done for exact plane waves, the two metric functions $P_1$ and $P_2$ would have been identical as there is no matter present and the wavefronts are planar. Hence, the metric solutions and the subsequent analysis of memory effects would be independent of $\kappa$.}

\noi The other components of the stress energy tensor can simply be computed from the relationships given in Eqs.(\ref{eq:vv_eqn_2}) and (\ref{eq:xx_eqn_2}).  The Ricci scalar curvature (physical metric) for such a solution becomes a constant, $R=2(\Lambda+\sigma)$. Depending on the signs of the cosmological constant and the flux parameter $\sigma$ we obtain solutions that have different background geometries ($S^2$ or $H^2$).

\subsection{Energy conditions and constraints}

\noi For the Null Energy Condition, the null vector is taken as $k^{\mu}=\delta_v\,^\mu$. We find that
$$T_{\mu\nu}k^{\mu}k^\nu=0.$$
\noi This shows there is no null flux present as matter. The Weak Energy Condition evaluated for a timelike vector ($t^\mu$) is shown below.
$$t^\mu=\frac{1}{\sqrt{2}}[\delta_u\,^\mu+(1-H/2)\delta_v\,^\mu] \hspace{2cm} T_{\mu\nu}t^{\mu}t^\nu=\sigma.$$ 
\noi This justifies why $\sigma$ is denoted as the flux parameter. Thus, the flux has to be positive-definite, $\sigma\geq0$. The second constraint follows from the square root in the metric function $P_1$ (see Eq.(\ref{eq:metric_P1_P2})).
$$1+\kappa(\Lambda+\sigma)\geq0$$
\noindent We will assume $\sigma=0$ in our entire geodesic analysis. Hence, there is no flux present perpendicular to transverse spatial wave surfaces.  Moreover, we find from the above constraint that $\alpha>0 \hspace{0.1cm} (<0)$ depending on whether the Ricci scalar $R>0 \hspace{0.1cm} (<0)$. Note that for a given background spacetime, $\alpha$ and $\kappa$ are related (Eq.(\ref{eq:P_2_eqn})). 

\subsection{Geodesic analysis of memory effect}

\noi The geodesic equations for the physical metric given in Eq.(\ref{eq:phys_metric_KW}) are
\begin{gather}
    \ddot{x}+\frac{P_1,_x}{P_1}(\dot{y}^2-\dot{x}^2)-2\dot{x}\dot{y}\frac{P_1,_y}{P_1}+\frac{1}{2}H_1,_x P_1\,^2=0 \label{eq:x_eqn_KW1} \\
     \ddot{y}+\frac{P_1,_y}{P_1}(\dot{x}^2-\dot{y}^2)-2\dot{x}\dot{y}\frac{P_1,_x}{P_1}+\frac{1}{2}H_1,_y P_1\,^2=0 \label{eq:y_eqn_KW1}
\end{gather}
\noi  The radiative term $H_1(u,x,y)=\sech^2(u) (x^2-y^2)$ is chosen to represent a {\em sech-squared pulse}.  Note that $\dot{x}=\dot{y}=0$ at $u\to -\infty$ can be taken as the initial condition.  We solve for different values of $\kappa$ with a  fixed $\Lambda$. The signature of $\Lambda$ decides the background geometry \citep{Chak1:2020}. We will consider both cases with positive and negative scalar curvature. 

\noi Eqs.(\ref{eq:x_eqn_KW1}) and (\ref{eq:y_eqn_KW1}) have earlier been solved numerically in \citep{Chak1:2020} for different functional forms of $P_1$ (see Eqns. (7), (8), (10) and (11)) in the context of GR. Here, we carry out a similar analysis with spacetimes having different values of $\kappa$. {\em Since $\alpha$ and $\kappa$ are related} (Eq.(\ref{eq:P_2_eqn})), {\em we will try to infer the behavior of  {geodesic separation} from the variation in $\alpha$.} 

\subsubsection{Negative scalar curvature}

\noi We solve geodesics with $\Lambda=-0.25, R=-0.5 \hspace{0.1cm}(\alpha<0)$. The evolution of $x$ and $y$ coordinates are  obtained by numerically solving the geodesic equations in {\em \footnote{For both the sources (generic matter, EM field) we have used we have used {\em Mathematica 10} for numerically solving the geodesic equations and the geodesic deviation equations, and also for obtaining the plots.} Mathematica 10}.

\begin{figure}[H]
    \centering
    \hspace*{-0.8cm}\includegraphics[scale=0.6]{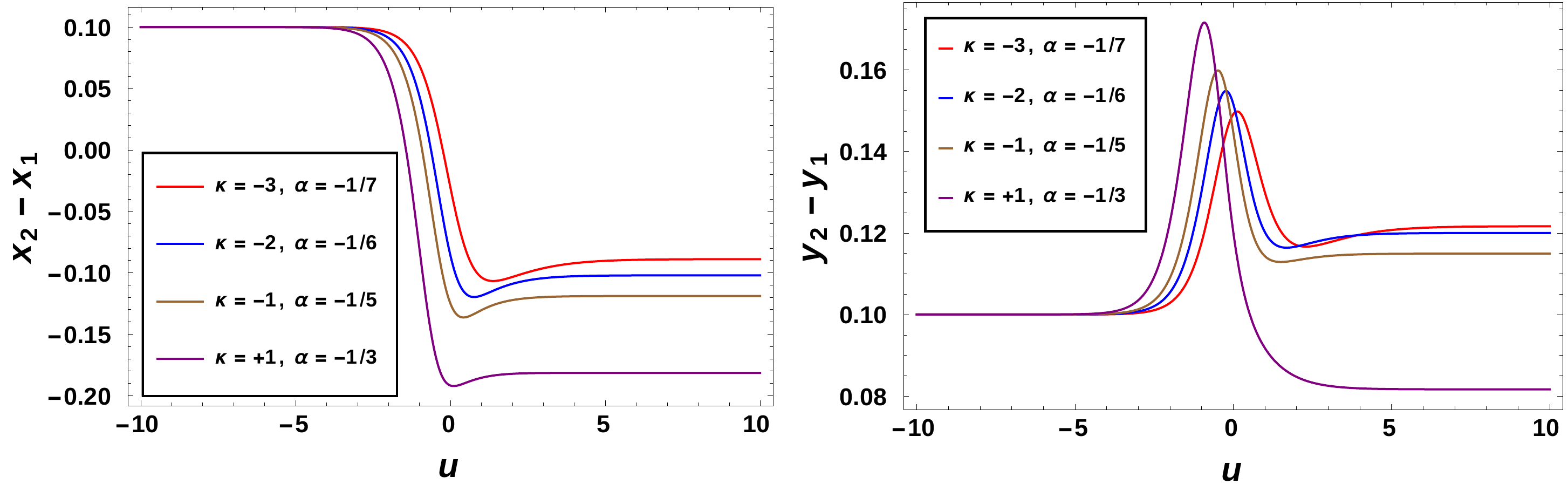}
    \caption{{Geodesic separation for negative scalar curvature along $x$ (left) and $y$ (right) directions having initial value $(0.1, 0.1)$, respectively.}}
    \label{fig:Neg_KW}
\end{figure}

\noi {In Fig.(\ref{fig:Neg_KW}), we plot geodesic separation, {\em i.e.} $x_2-x_1$ and $y_2-y_1$. Here,  $x_1,y_1$ and $x_2,y_2$ are solutions of a pair of geodesics along $x$ and $y$ directions, respectively, with different initial positions but zero relative velocity.}  {{\em One observes that the final value of geodesic separation is different from the initial value. This signifies the presence of gravitational wave memory.}} Along the $x$-direction we find that, as the negative value of $\alpha$ increases, the displacement of the separation from its initial value increases. For the $y$-direction plot, we observe a maxima near $u=0$ for all values of $\kappa$. The higher value of $\lvert \alpha \rvert$, the higher the peak of the maxima. But, in both plots of Fig.(\ref{fig:Neg_KW}), we observe that the separation is constant with zero relative velocity.  Thus, we find {\em constant shift displacement memory} in all these results. There is no velocity memory effect present. This result is similar with those found in GR \citep{Chak1:2020}.

\noi From the geodesic equations (\ref{eq:x_eqn_KW1}) and (\ref{eq:y_eqn_KW1}), it might appear that the behavior of the $x$ and $y$ coordinates should be identical. This is not observed in the plots due to the functional form of the gravitational wave term $H_1(u,x,y)$. We consider plus polarization and hence, $x$ and $y$ are not symmetric. Instead, if we had worked with cross polarization, the memory effects would have been identical along both directions.

\subsubsection{Positive scalar curvature}

\noi Geodesics in the positive scalar curvature spacetime are studied numerically with $\Lambda=0.25$ and $R=0.5 \hspace{0.1cm} (\alpha>0)$.

\begin{figure}[H]
 \hspace*{-0.8cm}	\includegraphics[scale=0.6]{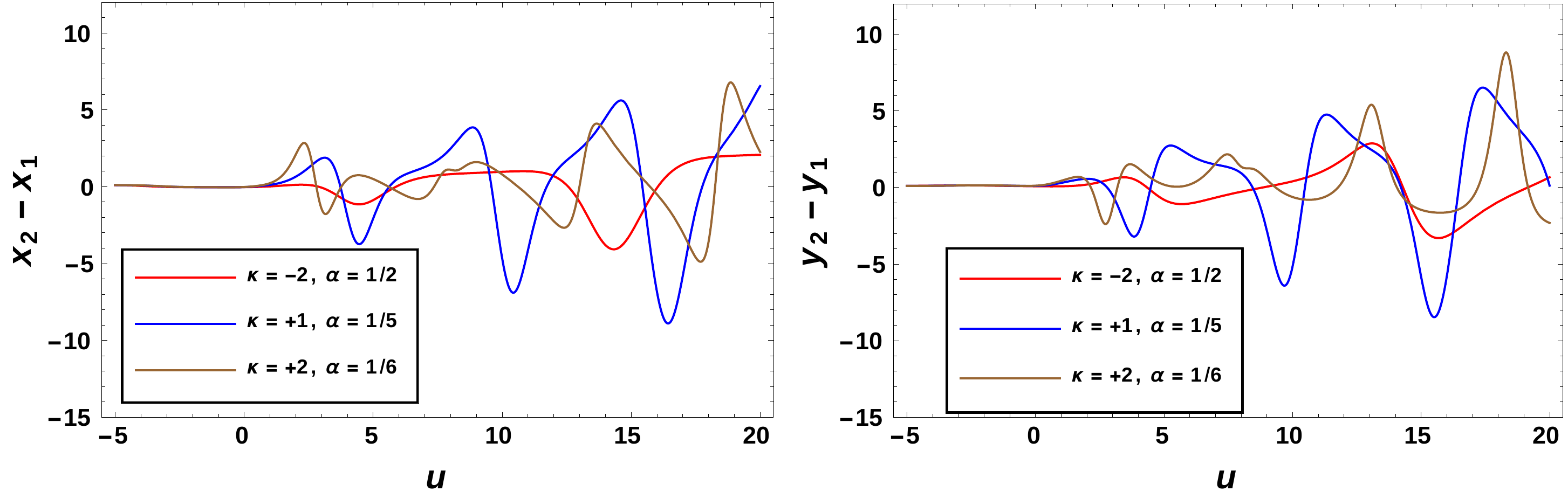}
\caption{{Geodesic separation for positive scalar curvature along $x$ (left) and $y$ (right) directions having initial value $(0.1, 0.1)$, respectively.}}
\label{fig:Pos_KW}
\end{figure}

\noi {Positive curvature solutions in GR were shown to give rise to a {\em frequency memory effect} \citep{Chak1:2020}. In that scenario, each geodesic was associated with a particular frequency. Thus, the geodesic separation lead to the formation of beats (see Fig.(3b) and (5b) in \citep{Chak1:2020}). In the present context, we find similar formation of beats where the central frequency is dependent on $\kappa$ as can be seen in Fig.(\ref{fig:Pos_KW}). Lower the value of $\alpha$, more is the frequency of oscillation along both the directions. }

\begin{comment}
\begin{figure}[H]
	\centering
	\begin{subfigure}[t]{0.4\textwidth}
		\centering
		\includegraphics[width=\textwidth]{sep_x_pos_const.eps}
		\caption{x direction}
		\label{fig:sep_x_pos_KW1}
\end{subfigure}\hspace{1.5cm}
	\begin{subfigure}[t]{0.4\textwidth}
		\centering
		\includegraphics[width=\textwidth]{sep_y_pos_const.eps}
		\caption{\centering y direction.}
		\label{fig:sep_y_pos_KW1}
	\end{subfigure}
	\caption{Separation between the geodesics for positive scalar curvature having initial separation as $\{(x, y)=(1,0.2)\}$}
	\label{fig:sep_pos_KW1}
\end{figure}

\noi The geodesic separation plots shown in Fig.(\ref{fig:sep_pos_KW1}) also demonstrate the feature of frequency memory. We have a beat formation due to the difference in the separation ({\em i.e.} difference in frequency) between a pair of geodesics. For more details on the nature of frequency memory effect, we refer the readers to \citep{Chak1:2020}. 
\end{comment}

\subsection{Geodesic deviation analysis}

\noi We now discuss memory effects obtained using the geodesic deviation equation. Using the procedure from our earlier work on Kundt waves in Brans-Dicke theory \citep{Siddhant:2020}, we construct the orthonormal tetrad from the physical metric given in Eq.(\ref{eq:phys_metric_KW}).
\begin{equation}
\begin{split}
 & e_0\,^{\mu}=[1,\dot{v},\dot{x},\dot{y}] \hspace{2.8cm} 
    e_1\,^{\mu}=\bigg[0,-\frac{\dot{x}}{P_1},-P_1,0\bigg] \\
 & e_2\,^{\mu}=\bigg[0,-\frac{\dot{y}}{P_1},0,-P_1\bigg]  \hspace{1.5cm} 
    e_3\,^{\mu}=[-1,1-\dot{v},-\dot{x},-\dot{y}] \label{eq:tetrad_KW_phys}
    \end{split}
\end{equation}

\noi The construction of a similar orthonormal tetrad as given in Eq.(\ref{eq:tetrad_KW_phys}) was earlier worked out in \citep{Bicak:1999}. We find that $ e_1\,^{\mu}$ and $ e_2\,^{\mu}$ are not parallelly transported and, hence, they are rotated by an angle $\theta_p$. 
\begin{equation}
    \dot{\theta_p}= \frac{1}{P_1}(P_1,_y\dot{x}-P_1,_x\dot{y}) \label{eq:parallel_transport}
\end{equation}

\noi The expressions for the Riemann tensor in this tetrad frame are provided in the Appendix. Using these expressions (Eqs.(\ref{eq:R1010_B})-(\ref{eq:R2020_W})) in the deviation equations (\ref{eq:deviation_eqn1_bg}) and (\ref{eq:deviation_eqn1_wave}), we numerically solve the geodesic deviation (background and wave separately) in the tetrad frame. We find that $Z_0$ and $Z_3$ have no evolution.\footnote{From the expressions of the Riemann tensors in the tetrad frame given in the Appendix, we observe that there are no terms like $R^0\,_{ijk}=R^3\,_{ijk}=0$. Hence, we set $Z^0=Z^3=0$ as they have no evolution.} Thus, we only evaluate the nontrivial behavior of $Z^1, Z^2$. Eventually, we go over to the coordinate basis using Eq.(\ref{eq:dev_tetrad}) and plot the background, wave and total deviation for different values of $\kappa$.
\begin{equation}
    \xi^x=-P_1Z^1 \hspace{2cm} \xi^y=-P_1Z^2 \label{eq:relation}
\end{equation}

\noi The deviation analysis is particularly useful as it separately gives the gravitational wave contribution from  the background. The total deviation, as shown earlier, is obtained by summing the contributions coming from the wave and the background. Throughout the text, we will clarify the similarity between the qualitative features of memory effects obtained from the deviation analysis and the geodesic analysis, whenever required.

\subsubsection{Negative curvature}

\noi We perform the geodesic deviation analysis for the same value of the cosmological constant ($\Lambda=-0.25$) and Ricci scalar ($R=-0.5$) as was used in the geodesic analysis. We assume in all the cases that the initial deviation value is $\xi^x=0.1,\xi^y=0.1$.

\begin{figure}[H]
    \centering
 \hspace*{-0.8cm}    \includegraphics[scale=0.6]{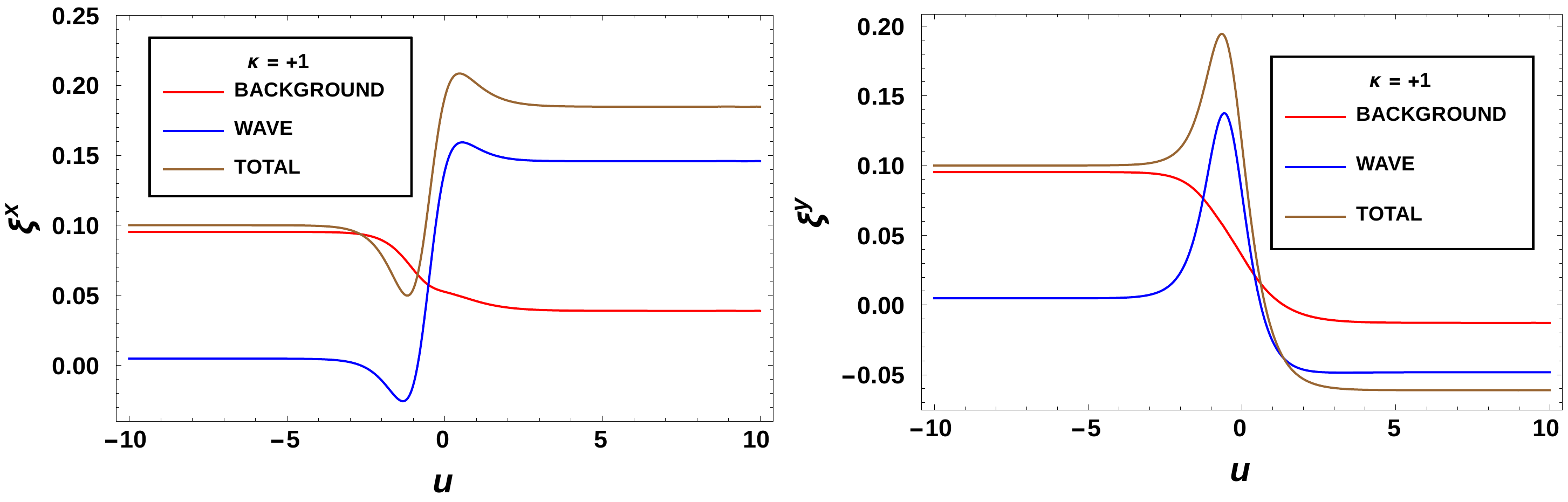}
    \caption{   {Coordinate deviation between geodesics along $x$ (left) and $y$ (right) directions having initial separation $\{(\xi^x, \xi^y)=(0.1,0.1)\}$ for $\kappa=+1$.}}
    \label{fig:dev_minus2_neg}
\end{figure}

\noi The plots in Fig.(\ref{fig:dev_minus2_neg}) show the background, wave and total deviation for $\kappa=+1$. In both cases, the background contribution decreases.  We find a rise in wave deviation along $x$-direction, while along the $y$-direction, it peaks around $u=0$ and then finally settles to a constant value. We find {\em constant shift displacement memory} along both the directions. The total deviation settles to a final value and, hence, no velocity memory is observed. 

\vspace{0.5in}

\begin{figure}[H]
    \centering
 \hspace*{-0.8cm}    \includegraphics[scale=0.6]{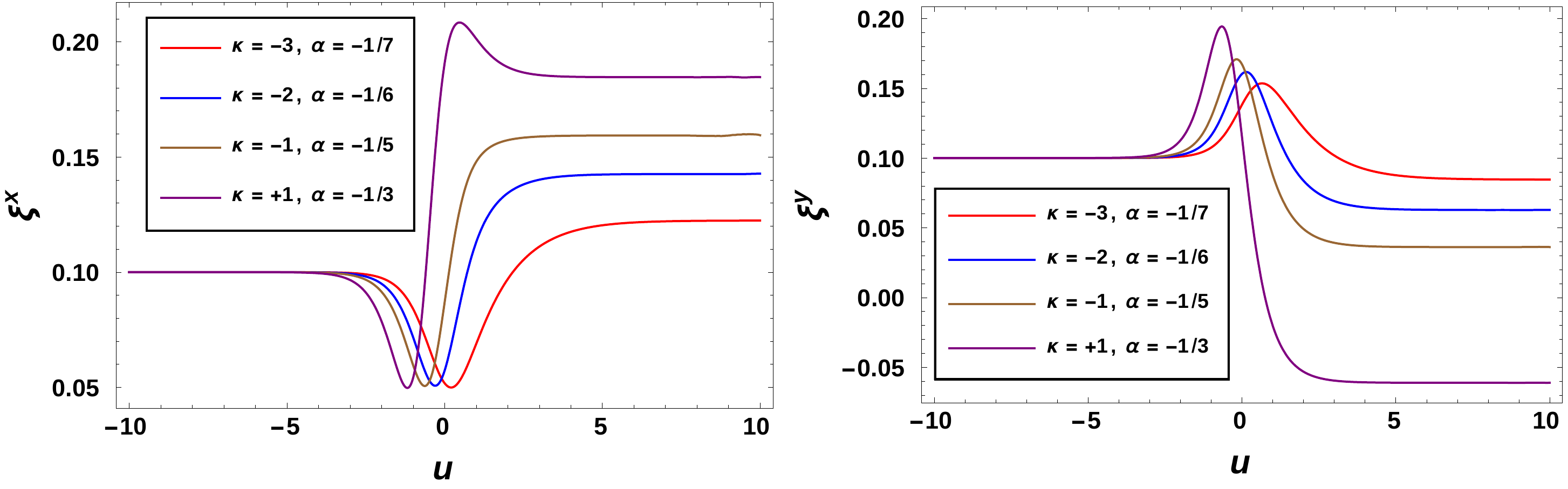}
    \caption{   {Total deviation between geodesics having constant negative scalar curvature along $x$ (left) and $y$ (right) directions.}}
    \label{fig:total_neg}
\end{figure}

\noi The plots in Fig.(\ref{fig:total_neg}) demonstrate the effect of EiBI parameter $\kappa$ on the total deviation. 
In the plot for the $x$-direction, we find that the total deviation rises with the increase in the negative value of $\alpha$.  In the $y$-direction plot, we find a maxima centered near $u=0$. The peak of the maxima rises with the rise in the absolute value of $\alpha$.  These results are consistent with the ones obtained from the geodesic analysis. The total deviation along $y$-direction also shows the same qualitative behavior with variation in $\alpha$, as obtained for the $x$-direction. Note that along both the directions we obtain constant shift displacement memory effect. 

\noi %In the geodesic analysis, we had shown the behavior of a single geodesic. Here, in the deviation part, we have plotted the evolution of the geodesic separation with the central geodesic being the one used in the geodesic analysis. Thus, there is an {\em intrinsic difference} in the two cases we have analysed. But, we find that, in both scenarios the memory effect is {\em qualitatively similar.}

\begin{figure}[H]
    \centering
    \hspace*{-0.8cm}     \includegraphics[scale=0.6]{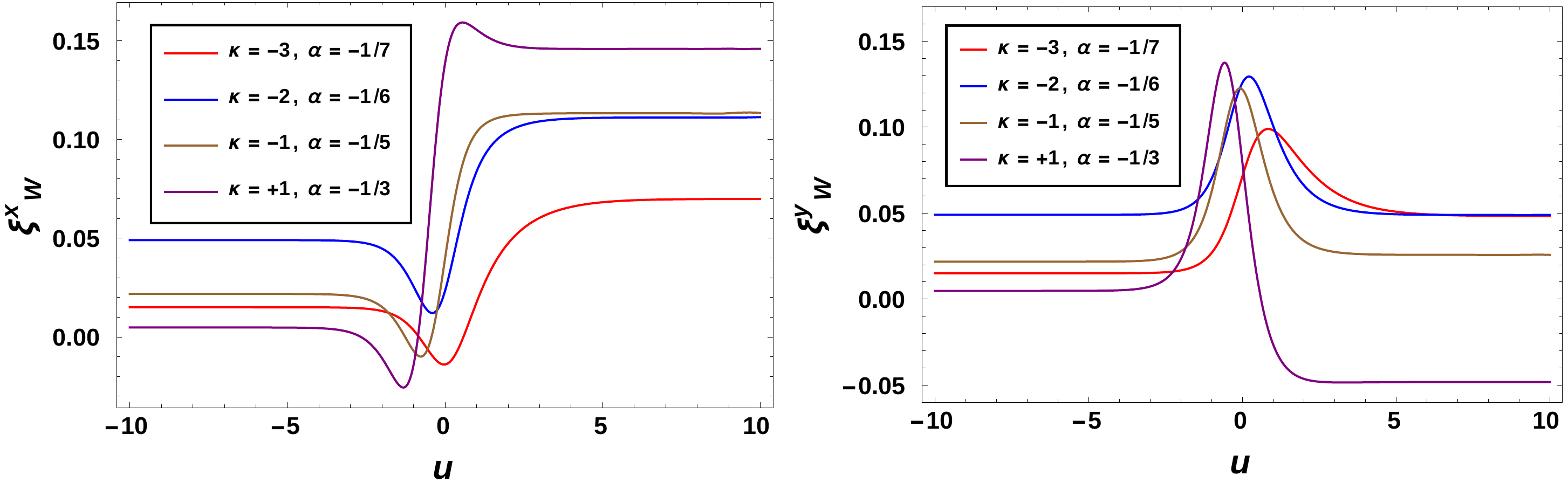}
\caption{   {Deviation due to gravitational wave between geodesics having constant negative scalar curvature along $x$ (left) and $y$ (right) directions.}}
    \label{fig:wave_neg}
\end{figure}

\noi The gravitational wave deviation ($\xi^x\,_W, \xi^y\,_W$) gives the actual measure of the {\em usual memory effect} used in the gravitational wave literature. We find that the variation of $\kappa$ (or $\alpha$) in the plots of Fig.(\ref{fig:wave_neg}) give similar results to total deviation.

\subsubsection{Positive curvature}

\noi The deviation analysis is done with the same value of $\Lambda=0.25$ and $R=0.5$ as used in the earlier geodesic analysis.  Here, we start with an initial separation, $\xi^x=0.1, \xi^y=0.1$. 

\begin{figure}[H]
    \centering
 \hspace*{-0.8cm}        \includegraphics[scale=0.6]{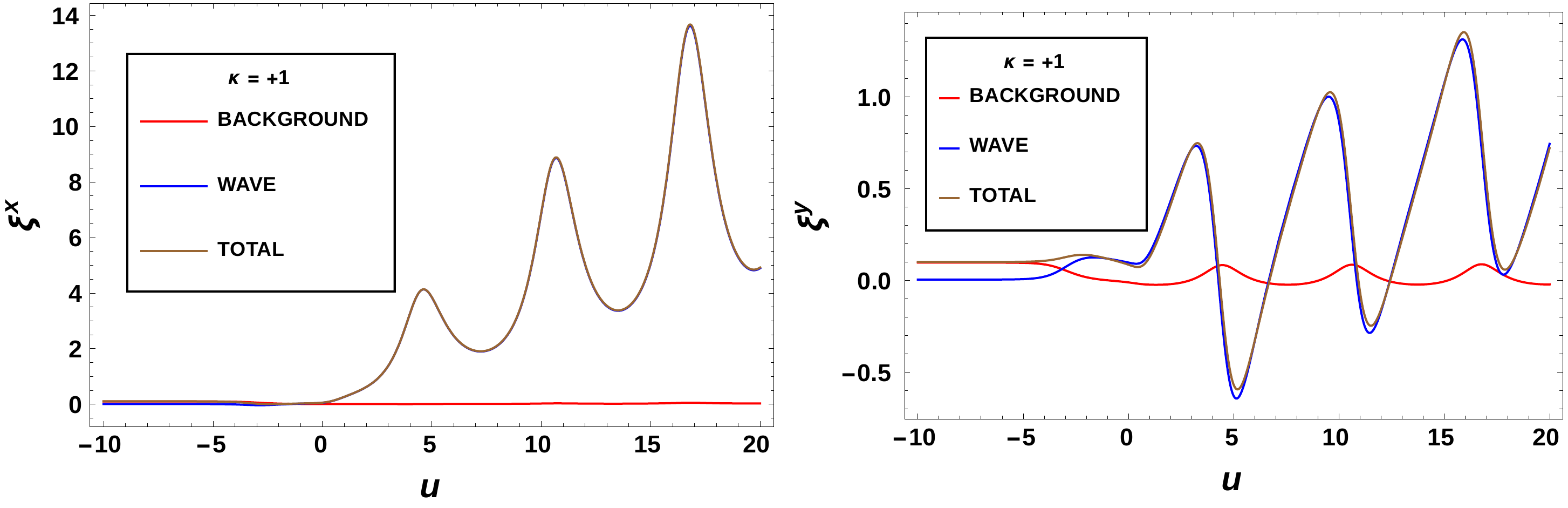}
    \caption{   {Coordinate deviation between geodesics having initial separation $\{(\xi^x, \xi^y)=(0.1,0.1)\}$ for $\kappa=+1$ along $x$ (left) and $y$ (right) directions.}}
    \label{fig:dev_minus2_pos}
\end{figure}

\noi Fig.(\ref{fig:dev_minus2_pos}) show the evolution of the deviation of the background, gravitational wave and their sum (total) along $x$ and $y$ directions respectively, for $\kappa=+1$.  We find that, in both cases, there is frequency memory effect and a subsequent beat formation. The background contribution is very small compared to the wave. Thus, the contribution to the total deviation comes mostly from the pulse of radiation present in the spacetime.

\begin{figure}[H]
    \centering
    \hspace*{-0.8cm}     \includegraphics[scale=0.6]{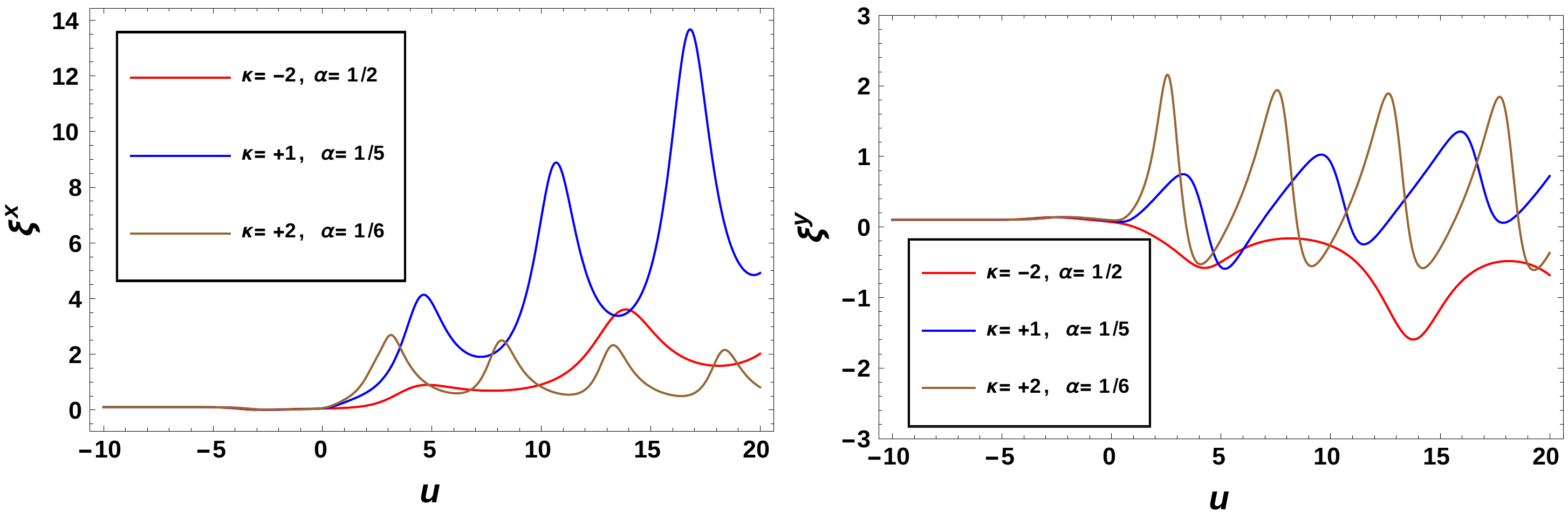}
    \caption{   {Total deviation between geodesics having constant positive  scalar curvature along $x$ (left) and $y$ (right) directions.}}
    \label{fig:total_pos}
\end{figure}

\noi The plots in Fig.(\ref{fig:total_pos}) give us the behavior of the total deviation for different values of $\kappa$. When the value of $\alpha$ decreases, we find that the frequency of the oscillation increases. In the geodesic analysis, we also found (from plots in Fig.(\ref{fig:Pos_KW})) similar behavior of frequency memory. Note that the geodesic and deviation plots are not identical in this scenario. This is due to the perturbative nature of the geodesic deviation equation. 

\begin{figure}[H]
    \centering
    \hspace*{-0.8cm}     \includegraphics[scale=0.6]{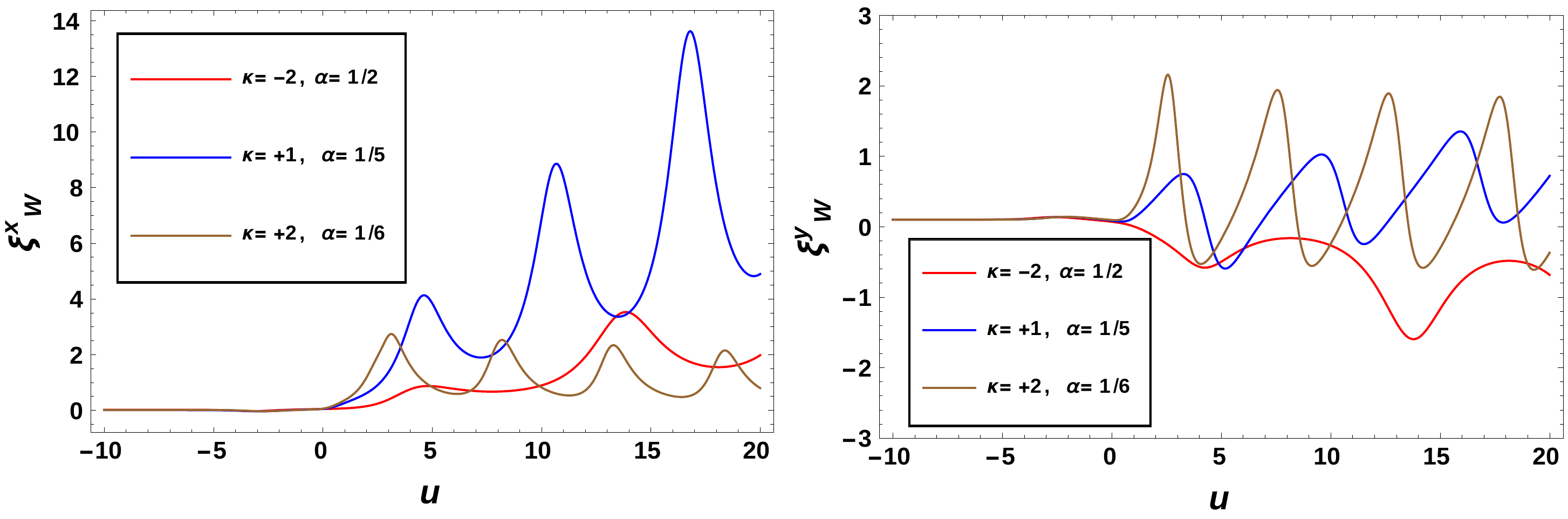}
    \caption{   {Deviation due to gravitational wave between geodesics having constant positive scalar curvature along $x$ (left) and $y$ (right) directions.}}
    \label{fig:wave_pos}
\end{figure}

\noi The deviation for the gravitational wave part in the plots of Fig.(\ref{fig:wave_pos}) is similar to the total deviation. This is because the total deviation gets most of its contributions from the gravitational wave pulse (see Fig.(\ref{fig:dev_minus2_pos})). 
\begin{comment}
\begin{figure}[H]
	\centering
	\begin{subfigure}[t]{0.4\textwidth}
		\centering
		\includegraphics[width=\textwidth]{geo_vs_dev_kminus2_x.eps}
		\caption{x direction}
		\label{fig:geo_vs_dev_x_pos}
\end{subfigure}\hspace{1.5cm}
	\begin{subfigure}[t]{0.4\textwidth}
		\centering
		\includegraphics[width=\textwidth]{geo_vs_dev_kminus2_y.eps}
		\caption{\centering y direction.}
		\label{fig:geo_vs_dev_y_pos}
	\end{subfigure}
	\caption{Comaprison between the total separation between geodesics obtained via the geodesic equation and the deviation equation for $\kappa=-2$.}
	\label{fig:geo_vs_dev_pos}
\end{figure}

\noi The comparison between the geodesic and the deviation plot in Fig.(\ref{fig:geo_vs_dev_pos}) highlights that initially they are similar but after the passage of the pulse the two results start to diverge. The deviation plot shows that with rise in $u$, the value increases rapidly than its geodesic counterpart. 
\end{comment}

\section{Kundt wave metric with an electromagnetic field source}

\noi Let us now turn towards the other exact solution. In solving the Kundt wave metric with an EM field, we do not {\em a priori} fix the functional form of the vector potential $A^\mu$. We start with a general electromagnetic field tensor $F_{\mu\nu}$ and solve the field equations of EiBI gravity. After obtaining the solutions, we look for $A_\mu$, which are consistent with our results. 

\subsection{Exact solution}

\noi We use the same ansatze given in Eqs.(\ref{eq:phys_metric_KW}) and (\ref{eq:aux_metric_KW}) for the physical and auxiliary metric, respectively. The energy momentum tensor for free Maxwell EM field is given as
\begin{equation}
    T^{\mu\nu}= F^\mu\,_\sigma F^{\nu\sigma}-\dfrac{1}{4}g^{\mu\nu}F_{\alpha\beta}F^{\alpha\beta}. \label{eq:Maxwell_EM}
\end{equation}

\noi First we will look into the field equation (\ref{eq:field_eqn_2}), since the equations for the other one will be similar ({\em i.e.} Eqs.(\ref{eq:uu_eqn_1}) and (\ref{eq:xx_eqn_1}) following from the field equation (\ref{eq:field_eqn_1})). The $uu$ component gives $T^{uu}=0$. Using this in Eq.(\ref{eq:Maxwell_EM}) yields
\begin{equation}
    F_{vx}=F_{vy}=0. \label{eq:F_vx}
\end{equation}

\noi The equations for components $ux, uy$ and $xy$ reduce to an identity.  For components $vx$ and $vy$ we find, respectively,
\begin{gather}
    F_{uv}F_{xu}=P_1\,^2F_{uy}F_{xy} \label{eq:vx_KW_2} \\
    F_{uv}F_{yu}=P_1\,^2F_{ux}F_{yx} \label{eq:vy_KW_2}.
\end{gather}

\noi Eqs.(\ref{eq:vx_KW_2}) and (\ref{eq:vy_KW_2}) simplify to give,
\begin{gather}
F_{ux}=F_{uy}=0. \label{eq:ux_uy_KW2}
\end{gather}
 
\noi The components $xx$ (or $yy$) yields a relation between the cosmological constant and  $F_{\mu\nu}$,
\begin{equation}
\Lambda=\frac{1}{2}(F_{uv}\,^2+P_1\,^4F_{xy}\,^2)     \label{eq:xx_KW_2}
\end{equation}

\noi We find that $\Lambda\geq 0$ from Eq.(\ref{eq:xx_KW_2}). Both the equations for components $uv$ and $vv$ yield
\begin{equation}
    \frac{1}{P_2\,^2}=\frac{1}{P_1\,^2}(1+2\kappa\Lambda) \label{eq:scaling_2}
\end{equation}

\noindent Thus, we again end up with a similar scaling relation between $P_1$ and $P_2$ like the one obtained in the earlier case (Eq.(\ref{eq:scaling})). Also, we find that $T^{uv}=\Lambda$. Thus, the constraint on \footnote{If we take the flux, $T^{uv}=0$, then the Ricci scalar vanishes. Hence we consider only positive values of $\Lambda$. }$\Lambda$ follows from the weak energy condition.  So, now, the differential equation satisfied by the metric function $P_2$ is given as
\begin{equation}
     P_2(P_2,_{xx}+P_2,_{yy})-P_2,_x^2-P_2,_y^2=\frac{2\Lambda}{1+2\kappa\Lambda} =\beta \label{eq:P2_KW_2}
\end{equation}

\noi Given the constraint on $\Lambda$, we find that the constant $\beta\geq0$.  Instead of using the former solution, we construct a new solution for $P_2$ from Eq.(\ref{eq:P2_KW_2}). We will find later how different choices of the function $P_2$ affects the nature of gravitational memory. Here, we assume that $P_2$ is independent of $y$, $P_2,_y=0$. The resulting ordinary differential equation has a solution like
\begin{equation}
  P_2(x)=\cosh(\sqrt{\beta}x)\hspace{1cm} P_1(x)=\sqrt{1+2\kappa\Lambda}\cosh(\sqrt{\beta}x) \label{eq:EM_soln}
\end{equation}

 \noi The Ricci scalar for the physical metric becomes $R=4\Lambda$. Therefore, we only have positive curvature solution ($\Lambda\geq 0$). For different sources we get different constraints on the metric solution.  We will try to understand how this constraint affects the gravitational wave memory in the spacetime.  

 \subsection{Maxwell field}
 
 \noi The field equations of EiBI gravity show that only components $F_{uv}$ and $F_{xy}$ are nonzero. But, they are constrained via the relation in Eq.(\ref{eq:xx_KW_2}). The Bianchi identity for the Maxwell equations yield
 \begin{equation}
     F_{uv}\equiv F_{uv}(u) \hspace{2cm} F_{xy}\equiv F_{xy}(x). \label{eq:Max_Bianchi}
 \end{equation}
 
\noi We assume the EM fields are independent of $v$ and $y$.  In order to have a dynamical electromagnetic tensor $F_{\mu\nu}$ we need to have a nonzero current $J^{\mu}$ sourcing the EM field.  The relevant gauge field $A_{\mu}$ and the source $J^{\mu}$, consistent with the results obtained above, are given below, along with the equations relating them.
\begin{gather}
    A_{\mu}=[0,A_v(u),0,A_y(x)]  \hspace{2cm} J^{\mu}=[0,J^v(u),0,0] \label{eq:Max_field_source}\\
    \frac{d^2 A_v}{du^2}=\frac{dF_{uv}}{du}=J^v \hspace{2cm} \frac{dA_y}{dx}= F_{xy}= \frac{B}{P_1^2} \label{eq:A_v_A_y}
\end{gather}
 
\noi Here, $B$ is a constant of integration.  Such a current source in Eq.(\ref{eq:Max_field_source}) satisfies the covariant conservation equation. Eq.(\ref{eq:xx_KW_2}) should always be satisfied by the EM fields. Having found the solution, we now investigate the memory effect. 

\subsection{Geodesic analysis of memory effect}

\noi We work out geodesic solutions for the same value of Ricci scalar as was done for the earlier used matter source, $R=0.5$. Hence, $\Lambda=0.125$ and $\beta>0$. The profile of the gravitational wave is taken as the sech-squared pulse [$h(u)=\sech^2(u)$].  In this scenario, as noted earlier, we only have positive curvature solution. The geodesic equations are same as that of Eqs.(\ref{eq:x_eqn_KW1}) and (\ref{eq:y_eqn_KW1}) with $P_1,_y=0$. Like the previous section, we will analyze the behavior of the {geodesic separation} w. r. t. the variation of $\beta$, since it is related to $\kappa$ via Eq.(\ref{eq:P2_KW_2}).

\begin{figure}[H]
    \centering
 \hspace*{-0.8cm}        \includegraphics[scale=0.6]{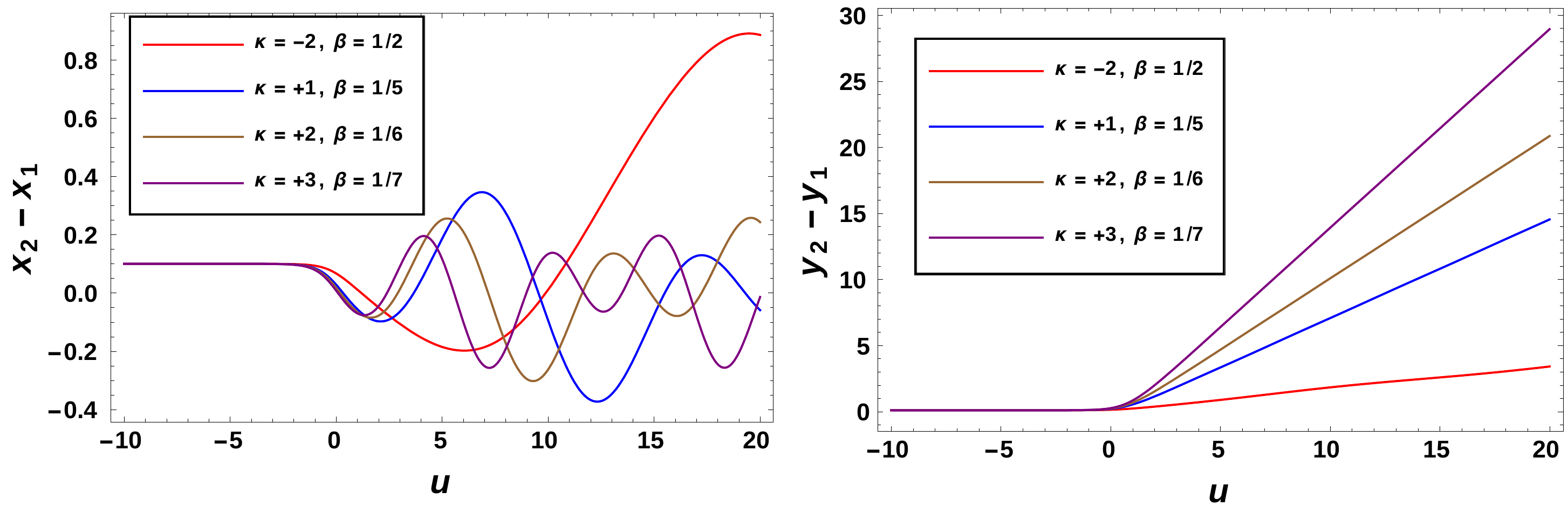}
    \caption{   {Geodesic separation along $x$ (left) and $y$ (right) directions with initial value as $\{(x,y)=(0.1,0.1)\}$.}}
    \label{fig:geo_EM}
\end{figure}

\noi Along the $x$-direction, the left plot in Fig.(\ref{fig:geo_EM}) shows the frequency memory effect. We find that the  frequency of oscillation is clearly dependent on $\kappa$ (or $\beta$). It decreases with the rise in the value of $\beta$. This was also seen for the earlier used matter source. We also find that the amplitude increases with the decline in the value of $\beta$. In the $y$-direction, we find a monotonically increasing displacement memory with decrease of $\beta$. Hence, there is presence of velocity memory effect. One should note that the values of $\beta$ used in the plots of Fig.(\ref{fig:geo_EM}) are identical with the values of $\alpha$ for the positive curvature solution of the previously used matter source. This happens because we examine the memory effects for the same value of Ricci scalar, in both the scenarios. 

\noi Another interesting observation is that, here,  we do not observe any frequency memory effect along the $y$-direction.  For the other matter source, we obtained frequency memory along both directions. Thus, this change in the behavior of memory effect is related to the functional dependence of $P_1$ on $x$ and $y$. In the former solution (Eq.(\ref{eq:metric_P1_P2})), $P_1$ was explicitly dependent on $x$ and $y$, while in the latter one (Eq.(\ref{eq:EM_soln})), it is only dependent on $x$.

\subsection{Geodesic deviation analysis}

\noi  The deviation analysis is done with the same orthonormal tetrad as given in Eq.(\ref{eq:tetrad_KW_phys}). The parallel transport condition (Eq.(\ref{eq:parallel_transport})) is also used for $e_1\,^\mu$ and $e_2\,^\mu$. We enlist the Riemann tensors in the tetrad frame in the Appendix.  Solving the required deviation equations in Fermi basis, we revert back to the coordinate basis. The results are obtained in terms of the following plots.

\begin{figure}[H]
    \centering
    \hspace*{-0.8cm}     \includegraphics[scale=0.6]{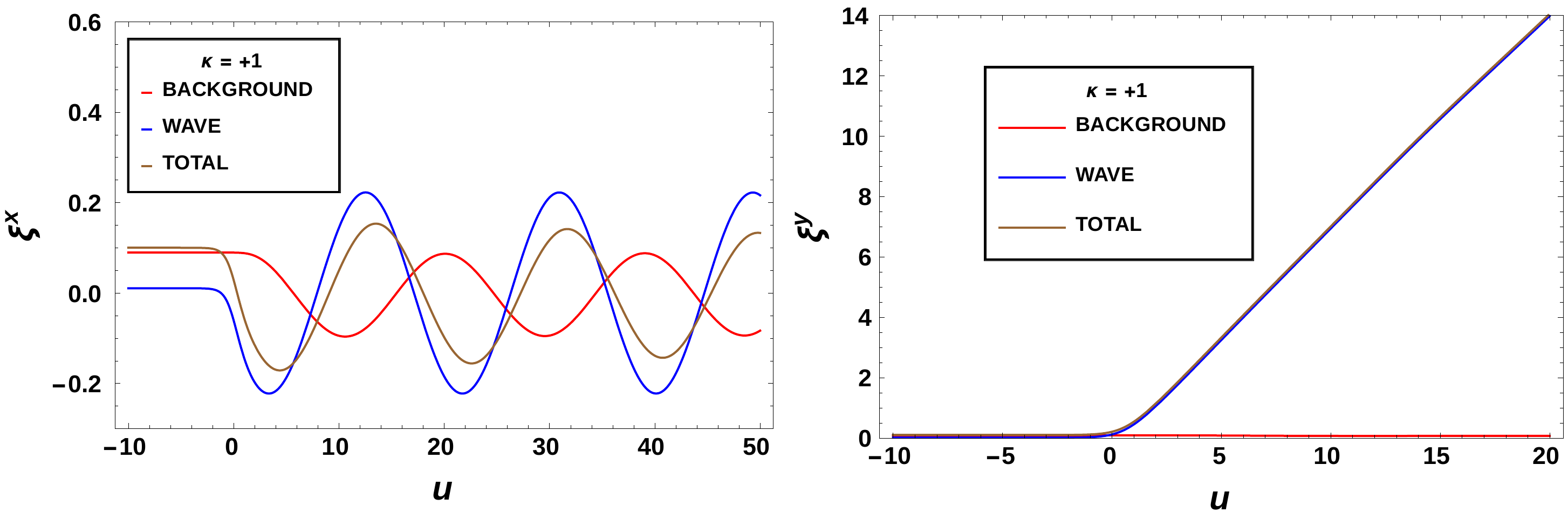}
    \caption{   {Coordinate deviation between geodesics having initial separation $\{(\xi^x, \xi^y)=(0.1,0.1)\}$ for $\kappa=+1$ along $x$ (left) and $y$ (right) directions.}}
    \label{fig:dev_k1_EM}
\end{figure}

\vspace{-0.5cm}
\noi In the plot for the $x$-direction in Fig.(\ref{fig:dev_k1_EM}), we again find the frequency memory effect for all the contributions. Thus, any non-radiative spacetime having negative background curvature (like AdS) can exhibit this oscillatory behavior of the geodesics.  We also find that the amplitude of wave deviation is higher than the background. Along the $y$-direction, we observe a monotonically increasing displacement memory. The entire contribution for the total deviation comes from the gravitational wave.

\begin{figure}[H]
    \centering
  \hspace*{-0.8cm} \includegraphics[scale=0.6]{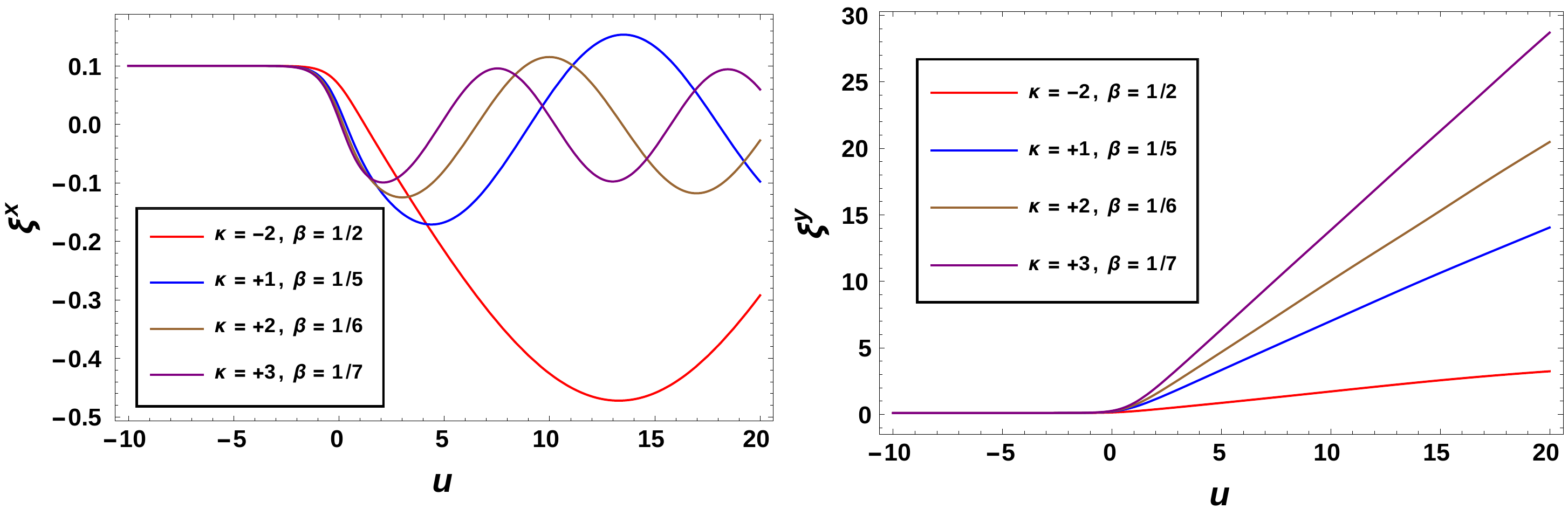}
    \caption{   {Total deviation in case of EM field along $x$ (left) and $y$ (right) directions.}}
    \label{fig:total_Em}
\end{figure}

\noi  The total deviation plot along $x$-direction in Fig.(\ref{fig:total_Em}) gives a similar result as obtained in Fig(\ref{fig:geo_EM}). The higher the value of $\beta$, the lower the frequency and higher the amplitude. Moreover, along the $y$-direction we find that with the rise in $\beta$, there is a decrease in the monotonic displacement memory of the total deviation. Thus, our results are completely in agreement with the geodesic analysis.

 \begin{figure}[H]
    \centering
 \hspace*{-0.8cm}        \includegraphics[scale=0.6]{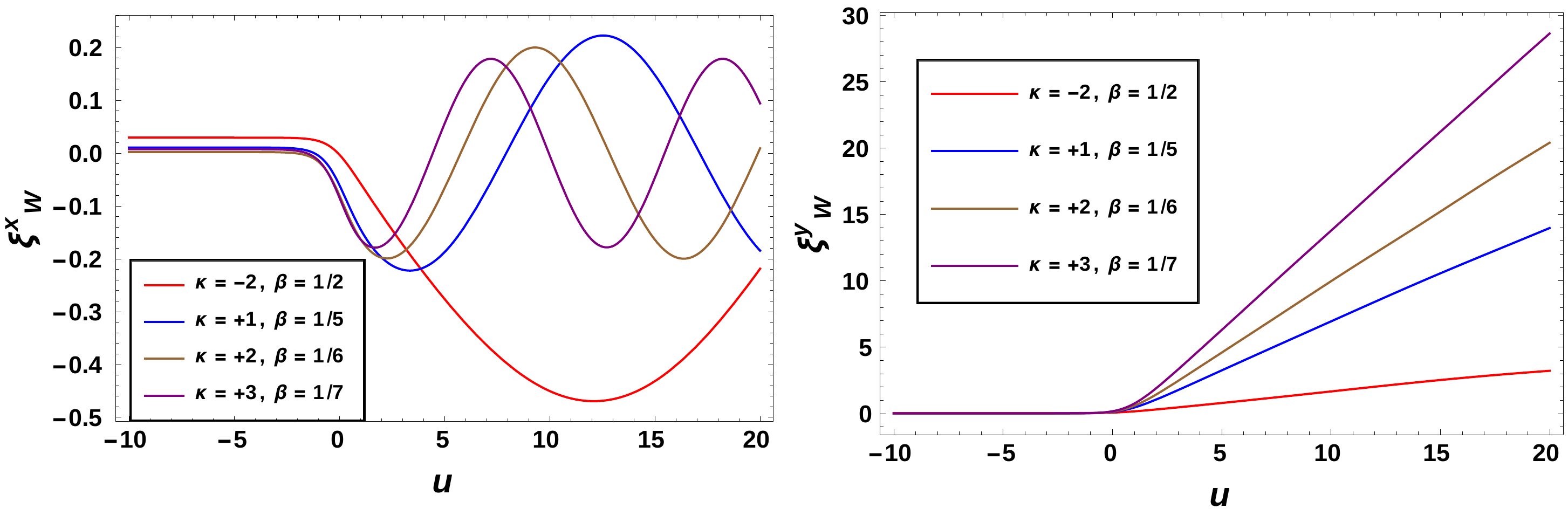}
    \caption{   {Deviation due to gravitational wave in case of EM field along $x$ (left) and $y$ (right) directions.}}
    \label{fig:memory_EM}
\end{figure}
 
\noi Fig.(\ref{fig:memory_EM}) gives the gravitational wave memory behavior with the variation of EiBI parameter $\kappa$. The plots are quite similar to the case of the total deviation. Along the $y$-direction, the behavior is identical as there is no contribution from the background. The entire deviation comes due to the memory effect.

\noi The above investigation on gravitational memory show that there is {\em no constant shift displacement memory for spacetimes sourced by EM fields.} This is because of the constraint on $\Lambda$, and thus, on the Ricci scalar. Hence, we infer that two different matter sources exhibit differences in the nature of memory effects.

\section{Conclusions}

\noi  {This article tries to explore, theoretically in the context of exact solutions, features of the gravitational wave memory effects in  EiBI theory of gravity.} To this end, we have first constructed novel solutions of Kundt wave geometries in this theory. Since EiBI is known to differ from GR at higher densities and curvatures, we have chosen geometries sourced by two different matter configurations. First, we solve for a generic matter source that satisfies all the energy conditions and field equations. Next, we find a solution for the electromagnetic field. Both the solutions exhibit an unconstrained function $h(u)$ which is responsible for determining the profile of the gravitational wave in the spacetime.  In both cases, we analyze memory effects by solving the geodesic equations and the geodesic deviation equations by choosing a {\em sech-squared pulse}. We observe that the matter content in the spacetime determines the nature of gravitational memory. Thus, the metric solution for the generic matter source acts as a tool to compare memory effects obtained for the EM field with itself.

\noi All the results in the paper are summarised in the following Table I. We write down the novel solutions obtained in EiBI gravity and the nature of memory effects they reveal. The change in gravitational wave memory corresponding to the variation in $\kappa$, via $\alpha$ (generic matter) and $\beta$ (EM source), is also presented.\begin{center}
\begin{table}[!h]
    {  \small
\hspace*{-2.2cm}\begin{tabular}{ |P{2.6cm}|P{3.3cm}|P{0.6cm}|P{3.2cm}|P{4.5cm}|P{5.2cm}|}
 \hline
 \multicolumn{6}{|c|}{GENERIC MATTER SOURCE} \\
 \hline
 \multirow{2}{*}{\textbf{Sign of scalar}}
& \multicolumn{3}{|c|}{\textbf{Metric functions}}  & \multicolumn{2}{|c|}{\textbf{Memory effect}}\\
\cline{2-6}
{\textbf {curvature} ($R$)} & $-g_{uu}=H_1$ &$g_{uv}$ &$g_{xx}^{-1/2}=g_{yy}^{-1/2}=P_1$  & Nature & Variation w.r.t $\alpha$ or $\beta$   \\
 \hline
$R<0,  \alpha<0$   & $\sech^2(u)(x^2-y^2)$&  -1& $\sqrt{1+\kappa\Lambda} \times \hspace{0.5cm} \bigg(1+\frac{\alpha}{4}(x^2+y^2)\bigg)$ & Constant shift displacement memory along both directions & Displacement memory increases with rise in $\lvert\alpha\rvert$ \\
\hline
$R>0, \alpha>0$ &  $\sech^2(u)(x^2-y^2)$ &-1& $\sqrt{1+\kappa\Lambda} \times \hspace{0.5cm} \bigg(1+\frac{\alpha}{4}(x^2+y^2)\bigg)$ & Frequency memory along both directions & Frequency of oscillation decreases with increase in  $\alpha$ \\
 \hline
 \multicolumn{6}{|c|}{ELECTROMAGNETIC SOURCE} \\
 \hline
  $R>0, \beta>0$ & $\sech^2(u)(x^2-y^2)$ & -1&$\sqrt{1+2\kappa\Lambda} \times \cosh(\sqrt{\beta}x)$ &Frequency memory only along $x$, displacement and velocity memory along $y$ &Frequency of oscillation decreases with increase in $\beta$, displacement and velocity memory decreases with rise in $\beta$  \\
 \hline
\end{tabular}\hspace*{-1cm}
 \label{Tab:results}
  \caption{Metric solutions and their corresponding memory effects for both the sources.}}
\end{table}
\end{center}

\vspace{-1cm}

\noi As EiBI gravity is a bimetric theory, we solve for both the physical and the auxiliary metric. We find that in case of both the sources, the gravitational wave part of the metric ($H_1$) are identical and is not dependent on the parameters of the theory, as can be seen from Table I. Moreover, the auxiliary metric turns out to be completely independent of $\kappa$. The induced spatial metrics (physical and auxiliary) on the $u$-constant wavefronts are conformally related where the conformal factor depends on $\kappa$. These wavefronts are curved because of the presence of matter. Hence, the background geometry (nonradiative part) for the physical spacetime is entirely theory dependent (through $\kappa$).  {This $\kappa$-coupling with the physical metric is not present in exact plane wave spacetimes as the background geometry is flat. This is the reason why we work with Kundt wave geometries.} 

\noi For the generic matter source we find a solution by choosing the flux parameter ($\sigma$) to be zero. We find that for nondynamical flux, the metric functions $P_1$ and $P_2$ are related via a $\kappa$-dependent scaling. The Ricci scalar turns out to be constant and depends on the cosmological constant ($\Lambda$). Hence, there is no restriction on the sign of the scalar curvature. 

\noi For the EM field, we do not {\em a priori} fix any form of the gauge field $A_{\mu}$. The field equations of the theory govern the behavior of the matter field. A consistent solution for the gauge field and the source current ($J^\mu$) is also provided. After solving the relevant field equations of EiBI gravity, we find an almost similar scaling relation between the metric functions $P_1$ and $P_2$ (Eq.(\ref{eq:scaling_2})) as was obtained earlier (Eq.(\ref{eq:scaling})). But here, the flux is equal to the cosmological constant. Thus, we only have constant non-negative Ricci scalar solutions.

\noi The constants $\alpha$ and $\beta$ are related to the EiBI parameter $\kappa$ via Eqs.(\ref{eq:P_2_eqn}) and (\ref{eq:P2_KW_2}), respectively. We use these constants to quantify the difference in the behavior of geodesic separation (from the geodesic analysis)  and the geodesic deviation for different values of $\kappa$. Both the analyses for the generic matter source show that in negative curvature spacetimes we observe constant shift displacement memory. The geodesic separation rises with the increase in the negative value of $\alpha$. This scenario is not permissible for the EM source as $\Lambda$ is strictly positive-definite. 

\noi In case of positive curvature spacetimes, we find that the behavior depends on the analytical forms of $P_1$. If $P_1$ is both $x,y$ dependent, then we find frequency memory along both  directions (as shown for the generic source). For $P_1\equiv P_1(x)$, we get frequency memory only along $x$. In the other direction, we get monotonically increasing displacement memory. The frequency of oscillation is found to decrease with the rise in the positive value of $\alpha$ (generic source) and $\beta$ (EM source). Hence, memory effects analyzed from geodesic and deviation analyses show logical consistency, as claimed earlier.

\noi   {Although qualitative features of memory effects as obtained here agree with GR \citep{Chak1:2020}, the EiBI parameter $\kappa$ produces an imprint on the amount of gravitational memory despite coupling only with the nonradiative part of the metric solution.} Also, studying this theory, one may find distinct memory effects based on the constraints imposed from the field equations, for different kinds of matter sources.

\noi A possible extension of this work can be done by introducing gyratonic terms in the Kundt wave line element and examining the memory effects in this new gyratonic Kundt spacetime. Also, one can study geodesic congruences for  such  Kundt  wave  geometries  and  calculate the ${\cal B}$-memory \citep{Loughlin:2019,Chakra:2020}. 

\noi Finally, we conclude by commenting that the link established  between the matter source and the nature of gravitational memory effects is worth investigating for diverse theories with nontrivial matter couplings.  {Even at the level of the exact solutions, such studies on Kundt geometries can generate different metric solutions corresponding to different gravitational theories. We hope to return to these issues in future.}

\section*{Acknowledgments}

\noi The author sincerely thanks Professor Sayan Kar for his constant motivation, for providing valuable insights during preparation of the manuscript and for carefully going through the final draft of the article. He also acknowledges Sayan Kumar Das and Soumya Chakrabarti for reading the manuscript. Finally, the author thanks the University Grants Commission (UGC), Government of India for providing financial assistance through a senior research
fellowship (SRF) (reference ID: 523711).

\bibliographystyle{apsrev4-2}
\bibliography{mybibliography_bd}

\section*{Appendix}

\noi The Riemann tensor components in the tetrad frame for the generic matter source are given below. 

{\em Background}
\begin{gather}
    (R^1\,_{010})_B=-\frac{1}{P_1^2}(\dot{y}\cos\theta_p+\dot{x}\sin\theta_p)^2[P_1,_x^2+P_1,_y^2-P_1(P_1,_{xx}+P_1,_{yy})] \label{eq:R1010_B}\\
      (R^1\,_{020})_B = \frac{1}{2P_1^2}[2\dot{x}\dot{y}\cos(2\theta_p)+(\dot{x}^2-\dot{y}^2)\sin(2\theta_p)][P_1,_x^2+P_1,_y^2-P_1(P_1,_{xx}+P_1,_{yy})]\\
      (R^2\,_{010})_B=\frac{1}{2P_1^2}[2\dot{x}\dot{y}\cos(2\theta_p)+(\dot{x}^2-\dot{y}^2)\sin(2\theta_p)][P_1,_x^2+P_1,_y^2-P_1(P_1,_{xx}+P_1,_{yy})]\\
        (R^2\,_{020})_B=-\frac{1}{P_1^2}(\dot{x}\cos\theta_p-\dot{y}\sin\theta_p)^2[P_1,_x^2+P_1,_y^2-P_1(P_1,_{xx}+P_1,_{yy})]
\end{gather}

{\em Gravitational wave}
 \begin{gather}
     (R^1\,_{010})_W= h(u)P_1[P_1\cos(2\theta_p)+(x\cos(2\theta_p)+y\sin(2\theta_p))P_1,_x+(y\cos(2\theta_p)-x\sin(2\theta_p))P_1,_y]\\
      (R^1\,_{020})_W= h(u) P_1[P_1\sin(2\theta_p)+(x\sin(2\theta_p)-y\cos(2\theta_p))P_1,_x+(x\cos(2\theta_p)+y\sin(2\theta_p))P_1,_y]\\
       (R^2\,_{010})_W=h(u) P_1[P_1\sin(2\theta_p)+(x\sin(2\theta_p)-y\cos(2\theta_p))P_1,_x+(x\cos(2\theta_p)+y\sin(2\theta_p))P_1,_y]\\
        (R^2\,_{020})_W=-h(u)P_1[P_1\cos(2\theta_p)+(x\cos(2\theta_p)+y\sin(2\theta_p))P_1,_x+(y\cos(2\theta_p)-x\sin(2\theta_p))P_1,_y]  \label{eq:R2020_W}
 \end{gather}

\noi In the case of the EM source, the above equations are valid with $P_1,_y=0$. Note that the analytical form of the function $P_1$ (see Eq.(\ref{eq:EM_soln})) is also different from the generic matter source.

\end{document}